# High-performance thin-film lithium niobate Mach−Zehnder modulator on thick silica buffering layer

Xiaotian Xue[1,2], Yingdong Xu[1,2], Wenjun Ding[1], Rui Ye[1], Jing Qiu[1,2], Guangzhen Li[1], Shijie Liu[1], Hao Li[1], Luqi Yuan[1], Bo Wang[1], Yuanlin Zheng[1,2,3,4,*] and Xianfeng Chen[1,2,3,4,5,*]

[1]State Key Laboratory of Advanced Optical Communication Systems and Networks, School of Physics and Astronomy, Shanghai Jiao Tong University, Shanghai 200240, China

[2]Zhangjiang Laboratory, Shanghai 201210, China

[3]Shanghai Research Center for Quantum Sciences, Shanghai 201315, China

[4]Hefei National Laboratory, 230088, Hefei, China

[5]Collaborative Innovation Center of Light Manipulations and Applications, Shandong Normal University, Jinan 250358, China

*Corresponding author e-mail: ylzheng@sjtu.edu.cn; xfchen@sjtu.edu.cn

**Abstract:** High-speed photonic integrated circuits leveraging the thin-film lithium niobate (TFLN) platform present a promising approach to address the burgeoning global data traffic demands. As a pivotal component, TFLN-based electro-optic (EO) Mach-Zehnder modulators (MZMs) should exhibit low driving voltage, broad operation bandwidth, high extinction ration, and low insertion loss. However, the pursuit of both maximal EO overlap integral and minimal microwave loss necessitates a fundamental compromise between driving voltage and operational bandwidth. Here, we demonstrate high-performance TFLN EO MZMs constructed on a 12-μm-thick silica buried layer using periodic capacitively loaded traveling-wave electrodes. In contrast to their counterparts utilizing undercut etched silicon substrates or quartz substrates, our devices exhibit streamlined fabrication processes and enhanced modulation efficiency. Notably, the fabricated MZMs attains a high modulation efficiency of 1.25 V·cm in the telecom C-band, while maintaining a low EO roll-off of 1.3 dB at 67 GHz. Our demonstration offers a pathway to achieving perfect group velocity matching and break the voltage-bandwidth limit in a simplified configuration suitable for volume fabrication, thereby laying foundational groundwork for the advancement of high-performance TFLN MZMs and benefiting the next-generation PICs in optical telecommunication, signal processing and other applications.

1. Introduction

Over the past two decades, the rapid progress of the information age, marked by the pervasive evolution of cutting-edge technologies such as 5G/6G networks, artificial intelligence (AI), ChatGPT, and cloud computing, has instigated an unprecedented surge in global data traffic[1,2]. This exponential growth poses challenges to data processing speed, operation bandwidth, and power consumption of signal processing devices[3], outstripping the capabilities of conventional electronic integrated circuits[4]. The next-generation ultra-high-speed photonic integrated circuits (PICs) based on the electro-optic (EO) effect present a promising solution to accommodate the high-velocity manipulation of massive data streams, enabling efficient EO signal conversion and signal processing[5,6]. EO modulators, which are the paramount component of such PICs, facilitate the mixing of microwave signals ($\sim 10^9$ Hz) and optical carrier signals ($\sim 10^{14}$ Hz) across different frequency domains, underpinning efficient data transmission, interconnection, processing, and computation capabilities[7,8]. The realization of ultra-high-performance EO modulators characterized by low driving voltage, broad bandwidth, low power consumption, negligible insertion loss, high extinction ratio (ER), and compatibility with large-scale manufacturing is essential for the future widespread implementation of PICs.

In the past decades, EO modulators based on various photonic platforms have been extensively investigated, including those based on silicon[7,9], lithium niobate (LN)[10–16], lithium tantalate (LT)[5], hybrid silicon nitride ($Si_3N_4$)[17–22], indium phosphide (InP)[23,24], polymers[25–27] and plasmonics[28–30]. Each photonic platform exhibits some unique performance merits, such as well-established fabrication processes (Si), exceptionally low propagation losses (hybrid $Si_3N_4$), capability for dense integration (Si, hybrid $Si_3N_4$, InP), high modulation efficiency (Si, InP, polymers, plasmonics), large bandwidth (LN, LT, polymers), high linearity (LN, LT)[31]. Nevertheless, achieving high-performance EO modulators simultaneously fulfilling all the desired performance metrics is still challenging.

Thin-film lithium niobate (TFLN) has recently emerged as a vigorous candidate for next-generation PICs with unparalleled performance. As a ferroelectric material, lithium niobate (LN) exhibits exceptional optical properties, such as an extensive transparent window (0.35 μm-5 μm), high Curie temperature, large EO coefficient (~30 pm/V), and strong nonlinearity (~-27 pm/V)[31,32]. The TFLN platform, combining the exceptional properties of LN with strong optical confinement, significantly enhances the EO interaction, thus leading to EO modulators with small device footprints and CMOS-compatible driving voltages[13,33–35]. The TFLN EO modulators, using coplanar waveguide (CPW) electrodes, have been experimentally demonstrated to have remarkable modulation efficiency of around 2.5 V·cm and a bandwidth ranging from 40 GHz to 67 GHz[10–12]. These advancements have fueled research into large-capacity telecommunication data transmission[8,12], EO frequency combs[36,37], and synthetic dimensions[38–40]. The driving voltage-bandwidth limit is inherently constrained by the intricate challenge of simultaneously optimizing for maximum EO overlap integral, group velocity matching, and microwave propagation loss. Recently, TFLN EO modulators based on micro-structured electrodes have garnered significant attention, as they effectively mitigate the microwave loss while maintaining high EO modulation efficiency. The innovation breaks the trade-off limitation between the driving voltage and the operational bandwidth, typically achieving modulation efficiencies of approximately 2.1 V·cm and 3-dB modulation bandwidths exceeding 100 GHz[15,19,41–45]. However, these modulators are typically constructed on partially etched silicon or quartz substrates to achieve optimal group velocity matching. Partially etched silicon substrates introduce both the design and fabrication complexities, which may be disadvantageous in volume fabrication. Alternatively, the lithium-niobate-on-quartz (LNOQ) approach necessitates a thicker upper cladding for precise microwave index adjustment, which reduces modulation efficiency and complicates the fabrication process, such as conductivity issues in electron beam lithography (EBL) process and necessity of the special end-face treatment.

In this work, we demonstrate a high-speed TFLN EO MZM using periodic capacitively loaded traveling-wave (CLTW) electrodes based on a thick silica buried layer of 12 μm on silicon substrate. Notably, the device features a high modulation efficiency of 1.25 V·cm in the telecom C-band due to a narrow T-rails gap and a thin upper cladding. Furthermore, we undertake an investigation into the relationship between the microwave index and electrode configuration, proposing an innovative strategy to achieve group velocity matching. This methodology offers valuable insights and serves as a pivotal reference for future research endeavors in developing high-performance EO modulators. Our device with a 7-mm-long interaction region showcases impressive performance metrics with a 1.3-dB roll-off in EO frequency response at 67 GHz. The work demonstrates a simplified configuration suitable for volume fabrication to achieving perfect group velocity matching and break the voltage-bandwidth limit for TFLN EO MZMs. Overall, our proposed TFLN EO modulator demonstrates high modulation efficiency and exceptional EO response, positioning it as a promising candidate for applications in the next-generation PICs.

## 2. Device design and fabrication

Figure 1(a) shows a schematic of the proposed TFLN EO MZMs using CLTW electrodes in a push-pull configuration based on thick silica buffering layer on silicon substrate. The detailed cross-sectional structure of the modulation region is depicted in Fig. 1(b). The device is designed and fabricated on a commercial x-cut TFLN wafer (NANOLN) composed of a 360-nm-thick LN layer and a distinct 12-μm-thick silica buried layer on a standard 500-μm-thick Si substrate, see also inset of Fig. 1(b). The fabrication process is described as follows: Firstly, the Mach-Zehnder interferometer pattern is defined using one-step electron beam lithography (EBL). Subsequently, the pattern is transferred to the TFLN layer by inductively coupled plasma (ICP) etching. The etching depth is 260 nm with the sidewall angle of about 60°. The top width of the ridge nanowaveguide in the modulation region is chosen to be 800 nm. Then, a two-step fabrication process is employed for the CLTW electrode. The T-rails electrodes, consisting of 600-nm-thick gold and 10-nm-thick Ti, are formed using a second EBL with polymethylmethacrylate (PMMA) and an electron beam evaporation (EBE) process for high-resolution patterning. The main electrodes, comprising 1.1-µm-thick gold and 20-nm-thick Ti, are patterned using UV lithography and deposited via a second EBE process (The CLTW electrode patterning can also be achieved using one-step DUV lithography). Next, the chip is coated with a 100-nm silica upper cladding layer by plasma-enhanced chemical vapor deposition (PECVD) for the purpose of group velocity matching. The silica buffer layer is then etched to expose the main electrodes and to ensure good conductivity at the contact with the RF probe. Finally, the end facets of the chip are finely polished to facilitate the light coupling. The fabrication processes are carried out at the Center for Advanced Electronic Materials and Devices (AEMD), Shanghai Jiao Tong University. Figure 1(c) presents a microscopy overview of the fabricated TFLN EO MZM employing CLTW electrodes. The light is coupled into/out of the TFLN MZM using a pair of lens fiber. The input light of $TE_0$ mode is evenly split into two arms of the MZI by a 50:50 multimode interferometer (MMI). The light then co-propagates with a microwave signal in a transmission line electrode and is modulated by the phase shifter using a push-pull configuration with a modulation length of 7 mm. The phase accelerated and retarded light in each MZI arm is recombined by another identical MMI connected to the output port, yielding an intensity modulated output.

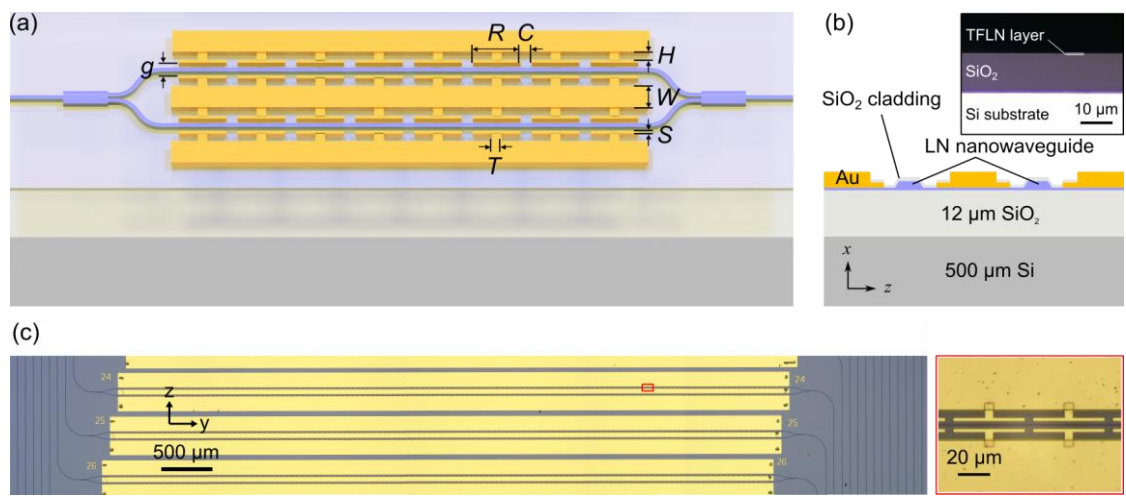

**Fig. 1. TFLN MZM on a thick silica buried layer.** (a) Schematic of the TFLN MZM.

In our optimal design, $W$, $g$, $H$, $S$, $T$, $R$, and $C$ are 50, 5, 5.2, 2, 5, 45, and 5 μm, respectively. (b) Cross section of the modulator (not to scale). Inset: Optical microscopy image of the cross section of the TFLN on 12-μm-thick $SiO_2$ buffering layer. (c) Optical microscopy image of the fabricated chip consisting TFLN MZMs employing the CLTW electrode. Inset: Image of the waveguide and the T-rails micro-structured electrodes.

While the group velocity matching condition can be readily obtained in TFLN based modulators compared to its bulk counterpart due to a low-relative-permittivity (~11.9) Si substrate, the microwave loss significantly increases. This loss arises from both metal loss, stemming from the finite resistivity of metals, and dielectric loss, resulting from the absorption by the buried layer and the substrate. As the electrode gap in TFLN modulators decreases to 5 μm or smaller, metal loss increases dramatically, yet this reduction in gap is necessary to enhance the modulation efficiency (half-wave voltage length product $V_\pi \cdot L$). To mitigate this issue, we incorporate a periodic CLTW electrode configuration, as shown in Fig. 1(c), featuring a CPW electrode with micro-structured T-rails extending from the main CPW into the gap. The T-rails effectively prevent the current from crowding on the electrode edges near the gap region, thereby significantly reducing the microwave loss without compromising modulation efficiency.

Notably, the microwave index of the CLTW electrode exhibits a substantial increase compared to that of the CPW electrode. The T-rails of the CLTW electrode essentially increase the capacitance per unit length of the electrode without inducing significant inductance alterations, causing the microwave refractive index to increase through the slow-wave effect. Previous research has leveraged quartz substrates with a low relative dielectric constant (~4) or partially etched silicon substrates to mitigate the high microwave index and achieve optimal velocity matching[41,42,44,45]. However, these approaches further introduce several design and fabrication challenges. For instance, utilizing quartz substrates can result in a microwave index much lower than the group index of light, necessitating a relatively thicker silica cladding to attain perfect velocity matching. Furthermore, the optical waveguide based on LNOQ features higher optical loss (lower thermal conductivity of substrate may lead to poorer etching quality), fabrication difficulty and yield issues, which is detrimental to large-scale integration or heterogeneous integration with silicon photonics. The fabrication process of TFLN on partially etched silicon substrate is also problematic due to precise requirements for etching depth and alignment between layers.

We take another approach and propose using TFLN with a 12-μm-thick silica buried layer, which can effectively reduce the microwave refractive index to a relatively high level. Consequently, a relatively thin cladding layer is sufficient to achieve perfect group velocity matching, facilitating a straightforward fabrication process compared to modulators based on LNOQ and TFLN with undercut etched silicon substrates. Notably, a thinner silica cladding layer enables a reduction in the half-wave voltage length product. This reduction can be attributed to the fact that the electric field at the interfaces between silicon dioxide and LN is smaller than that at the boundaries between air and LN, leading to a stronger EO overlap.

To validate the feasibility of achieving velocity matching using CLTW electrode on TFLN with a 12-μm-thick silica buried layer, we conduct simulations using finite element methods (Ansys HFSS) to assess the microwave index of various electrode configurations: CPW electrode on TFLN with silica buried layers of 4.7 μm, 12 μm, CLTW electrode on TFLN with a 12-μm-thick silica buried layer, and CLTW electrode on LNOQ. The model parameters for both CPW and CLTW electrode are sourced from relevant references[15,17]. All models are covered with a silica cladding layer of 100 nm thickness. The simulation results, presented in Fig. 2(a), demonstrate that the difference in microwave index between CLTW and CPW electrode can be compensated by adjusting the silica buried layer thickness from 4.7 μm to 12 μm. Furthermore, CLTW electrode based on TFLN with a 12-μm-thick silica buried layer exhibit a greater ease in achieving group velocity matching compared to CLTW electrode on LNOQ. The selected thickness of the silica buried layer is a preferred choice. On the one hand, when compared to the 4.7-μm (or thinner) silica buried layer, the 12-μm-thick buried layer significantly reduces the dielectric losses, thereby optimizing the response of the electrode, at the low-frequency range. On the other hand, the CLTW electrode exhibits a substantial reduction in ohmic losses, leading to a notable optimization for the high-frequency response of the electrode. Compared to LNOQ, our thinner silica upper cladding suffices to achieve velocity matching condition, ensuring a higher modulation efficiency.

Additionally, the microwave index positively correlates with the capacitance per unit length, pointing out a method for fine-tuning the microwave index. By adjusting the structure parameters of the T-rails, the capacitance of the CLTW electrode can be adjusted to achieve a precise microwave index necessary for perfect velocity matching. Figure 2(b) illustrates the microwave refractive index of the CLTW electrode when the parameter $H$ is equal to 1, 3 and 5 μm, respectively. Finally, combining with the simulation design, parameters of the CLTW electrode are derived as ($W$, $g$, $H$, $S$, $T$, $R$, $C$) = (50, 5, 5.2, 2, 5, 45, 5) μm. The corresponding simulation results for characteristic impedance and microwave index are presented in Figs. 2(c) and 2(d), respectively, indicating tolerable impedance matching and perfect velocity matching. Moreover, we simulated the optical field distribution of the transverse electric mode and the electrostatic field of the microwave electrodes, as shown in Fig. 2(e). The group index of the light is calculated to be 2.26 and a modulation efficiency of 1.2 V·cm is calculated at the wavelength of 1550 nm[31].

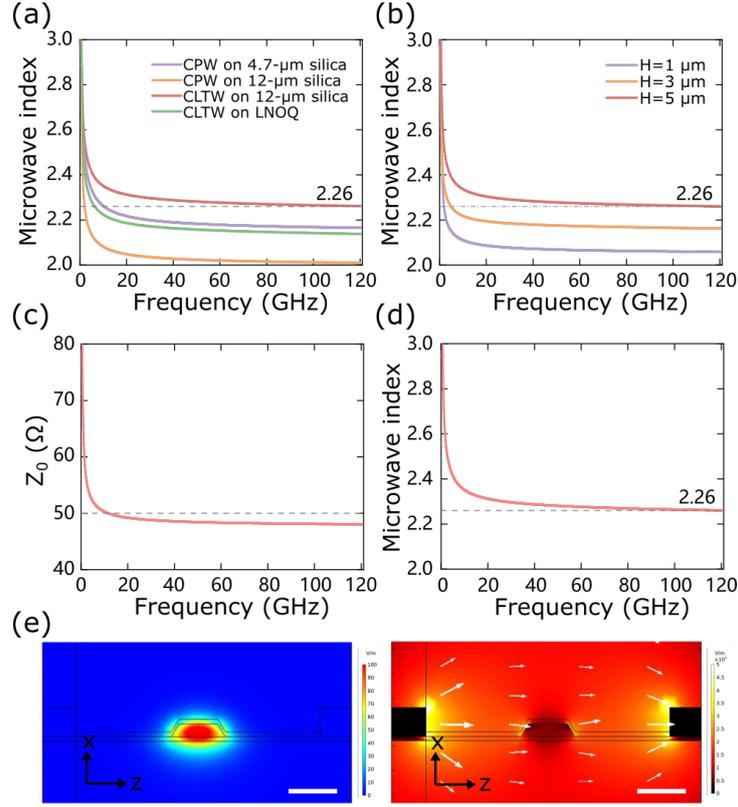

**Fig. 2.** (a) Simulated microwave indices under different configurations of electrodes based on different buried layer or substrate. (b) Simulated microwave indices for different $H$ parameters. (c) Calculated characteristic impedance $Z_0$. (d) Microwave index of the designed CLTW electrode. (e) Simulated optical and electrical field distribution. Scaler bar: 1 μm.

## 3. Measurement and Discussion

The photograph of the measurement platform and the modulators is presented in Fig. 3(a). Initially, the on-chip insertion loss of the MZM, encompassing both coupling loss and propagation loss, is assessed to be approximately 1.0 dB for the 7-mm device. The Fabry-Perot (F-P) interferometric method is utilized to evaluate the average optical loss of a phase modulator at 1550 nm. The waveguide propagation loss is measured to be approximately 0.3 dB/cm for a referenced waveguide with a total length of 17 mm on the same chip, which is similar to the loss without metal electrodes.

The static performance of modulation efficiency, denoted as $V_\pi \cdot L$, is measured at the wavelength of 1550 nm for modulation lengths of 4 mm and 7 mm. The $V_\pi$ values for the 4-mm and 7-mm modulators, as depicted in Figs. 3(b) and 3(c), are measured to be 3.14 V and 1.76 V, respectively. The measured $V_\pi \cdot L$ is approximately 1.25 V·cm, which aligns well with theoretical prediction. Compared to others based on CLTW electrodes listed in Table 1, our device exhibits significantly improved modulation efficiency. It is also feasible to achieve an ultra-low driving voltage of sub-1 V by narrowing the electrode gap or extending the length of the modulation region. Additionally, the extinction ratios, measured using a static voltage sweep, are approximately 30 dB. This is relatively high due to optimized fabrication processes, including improved write-field stitching compared to those based on LNOQ and precise electrode alignment facilitated by the use of EBL for electrode patterning.

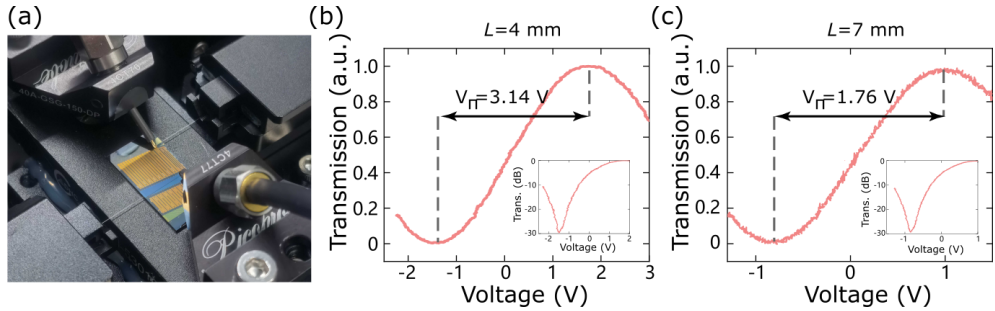

**Fig. 3.** (a) The TFLN MZM chip sample under test. Normalized optical transmission of the (b) 4- and (c) 7-mm-long modulators as a function of the applied voltage, showing a half-wave voltage of 3.14 V and 1.76 V, respectively.

Subsequently, we characterize the electrical-electrical (EE) $S$ parameters of the CLTW electrode, as depicted in Fig. 4(a). The measured EE $S_{21}$ displays a roll-off of -4.6 dB at 67 GHz for the device with a modulation of 7 mm. The EE reflection represented by $S_{11}$ remains below -22 dB across the entire measured frequency range. The characteristic impedance $Z_0$ approximately 46 Ω is extracted from the EE response as plotted in Fig. 4(b), showing a minor impedance mismatch. Notably, the extracted characteristic impedance encompasses contributions from all output ports, microwave cables, and RF probes, each featuring a 50-Ω characteristic impedance, suggesting that the intrinsic $Z_0$ of the CLTW electrodes might be less than 46 Ω. Figure 4(c) displays the extracted microwave index $n_m$ for the CLTW electrode. The microwave index in the high-frequency range closely aligns with the group index of the light (~2.26), reaching excellent velocity matching.

To substantiate the reduction in microwave loss afforded by the CLTW electrode, we conduct a comparison with a CPW electrode, under identical fabrication processes and gap dimensions. The fitted microwave losses are 0.74 dBcm$^{-1}$GHz$^{-1}$ for the CLTW electrode, and 1.05 dBcm$^{-1}$GHz$^{-1}$ for the CPW electrode, as illustrated in Fig. 4(d). The incorporation of T-rails diminishes microwave losses to approximately 70% of those in the CPW electrode. The microwave losses surpass the average values reported in the literature, and our device holds promise for achieving ultra-high performance.

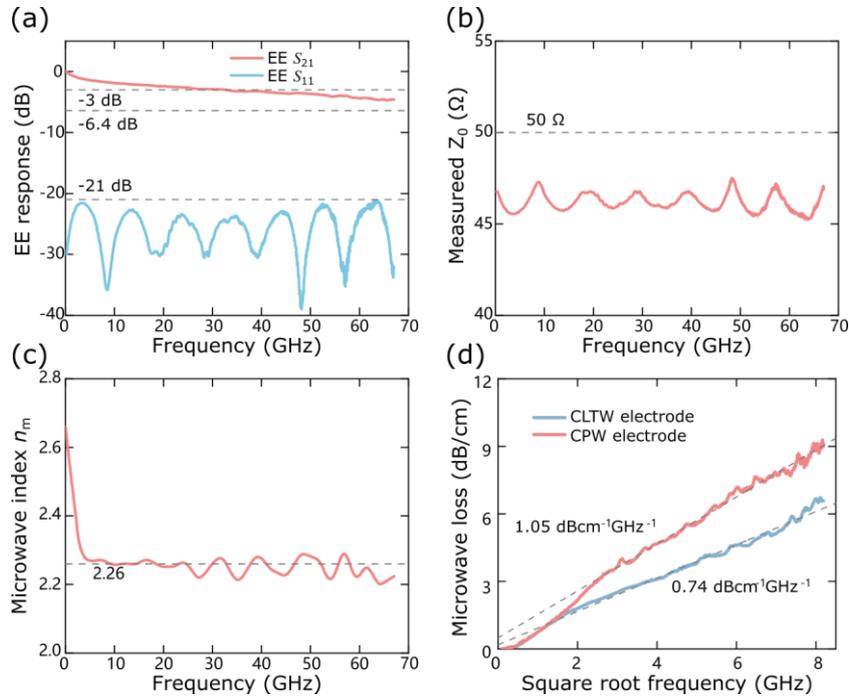

**Fig. 4.** (a) Measured EE coefficients of $S_{21}$ and $S_{11}$ for the 7-mm device, indicating a 3 dB bandwidth of approximately 35 GHz and a 6 dB bandwidth exceeding 67 GHz. Extracted (b) characteristic impedance $Z_0$ and (c) microwave index $n_\mathrm{m}$ for the CLTW electrode. (d) Measured microwave losses for both CLTW electrode and CPW electrode.

The small signal EO responses are gauged using a vector network analyzer (Agilent, N4373E). The RF signal from the VNA is applied to the electrode via a high-speed RF probe, while the other end of the electrode is connected to a 50-Ω termination through another probe. As shown in Fig. 5(a), the 3-dB EO bandwidth of the 7 mm device exceeds 67 GHz (limited by the VNA). The measured EO $S_{21}$ exhibits a rapid roll-off of approximately 1 dB at low frequency range below 2 GHz. This is attributed to the impedance mismatch, which is also evident from resonance ripples in the measured EO $S_{21}$ and $S_{11}$ response and the extracted characteristic impedance. The EO $S_{21}$ displays a flat roll-off from 2 GH to 67 GHz, with a roll-off as low as 1.3 dB at 67 GHz. The flat EO response in the high-frequency range reflects perfect group velocity matching. It should be noted that the tested frequency range is only limited due to the experimental constrains. The measured EO $S_{21}$ curve also shows the modulator has an extrapolated 3-dB bandwidth exceeding 120 GHz.

Furthermore, we have investigated a method to finely regulate the microwave index to achieve group velocity matching based on the CLTW electrode, providing a design model for high-performance EO modulators. The $H$ parameter, as illustrated in Fig. 1(a), affects the capacitance of the CLTW electrode and allows for adjustment of the microwave index. We extract and compare the microwave index for $H$ values of 1, 3, and 5 μm, as shown in Fig. 5(b). Additionally, the averaged microwave index values are extracted at high frequencies and plotted their relationship for different $H$ values in Fig. 5(c). The microwave index exhibits a linear relationship with $H$, with a fitted slope of 0.05 μm$^{-1}$. By adjusting the parameters of the T-rails, group velocity matching can be achieved.

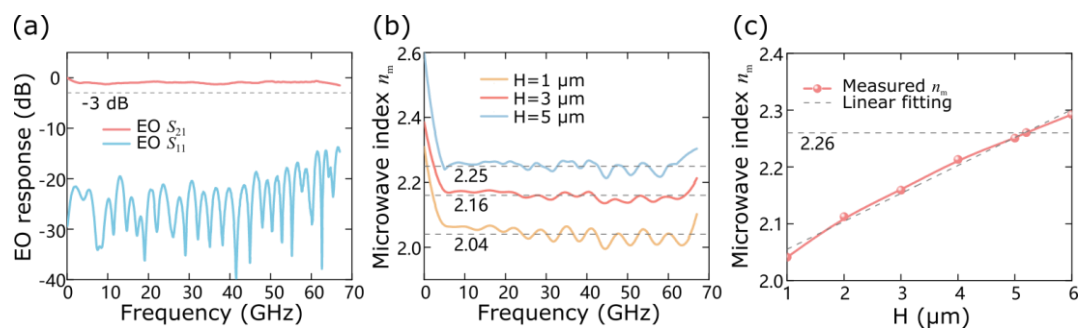

**Fig. 5.** (a) Measured EO responses $S_{21}$ and electrical reflection $S_{11}$ of the fabricated 7-mm-long modulator, showing a high 3-dB bandwidth significantly surpassing 67 GHz. (b) Calculated microwave index values for various $H$ parameters. (c) Liner correlation between microwave index values and different $H$ parameters.

To further demonstrate the outstanding performance of the fabricated 7-mm modulator, a detailed analysis of high-speed data transmissions is conducted, as depicted in Fig. 6. During the experiment, an arbitrary waveform generator (AWG) with an analog bandwidth of 65 GHz (Keysight 8199A) is employed to generate the RF signal, which is subsequently directly applied to the modulator. The input optical signal of 0 dBm is utilized, and the modulated output light is amplified by an erbium-doped fiber amplifier (EDFA) and refined by a band-pass filter. The output light is then detected by a high-speed photodetector and an oscilloscope (Keysight N1000A) with a bandwidth of 65 GHz is used to record the eye diagrams. Firstly, the OOK modulation is characterized at data rates of 80 Gb/s and 112 Gb/s, as correspondingly depicted in Figs. 6(a) and 6(b). Clear eye opening has been demonstrated. The measured dynamic ERs are found to be 7.14 and 4.42 dB, respectively. Additionally, PAM-4 modulation is implemented at 40 Gbaud (80 Gb/s) and 50 Gbaud (100 Gb/s), with the results presented in Figs. 6(c) and 6(d). The dynamic ERs reach 5.05 and 5.09 dB, respectively. Figure 6(e) portrays the back-to-back (B2B) bit error rate (BER) versus the received optical power for both 64 Gb/s OOK and 80 Gb/s PAM-4 signal transmissions. Notably, the BERs fall below the KP-4 forward error correction (FEC) threshold of $2.4\times10^{-4}$ in both scenarios. Specifically, under the PAM-4 signal transmission at 80 Gb/s, the BER can be further decreased to below the hard decision forward error coding (HD-FEC) threshold of $3.8\times10^{-3}$ when the received optical power exceeds -17 dBm. As the received optical power increases prior to pre-amplification, the BERs decrease linearly, with no observable error floor within the range of optical powers tested. Collectively, the results presented in Fig. 6 underscore the excellent linearity of the modulator for high baud-rate optical transmissions. It is pertinent to emphasize that the observed performance is likely constrained by factors such as the electronic digital-to-analog converter, the RF probe, and driver noise, rather than distortions inherent to the modulator itself.

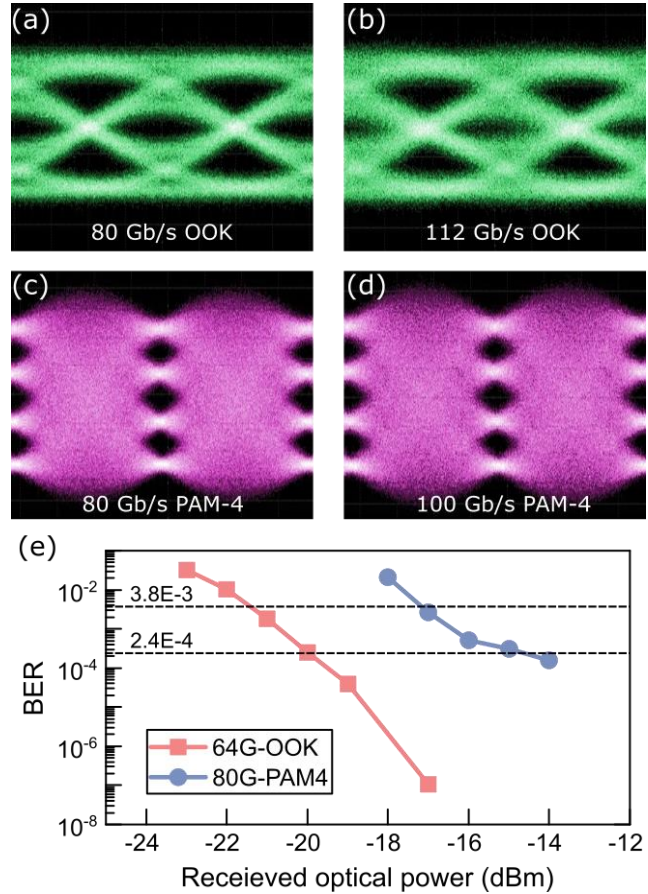

**Fig. 6. High-speed data transmission of the fabricated 7-mm modulator.** Measured optical eye diagrams for the OOK format at data rates of (a) 80 Gb/s and (b) 112 Gb/s. Measured optical eye diagrams for the PAM-4 format at data rates of (c) 40 Gbaud (80 Gb/s) and (d) 50 Gbaud (100 Gb/s). (e) The BER versus the received optical power for both 64 Gb/s OOK and 80 Gb/s PAM-4 signal transmissions.

Finally, we summarize the performance metrics of recent research on TFLN EO modulators in Table 1. Our device exhibits a lower $V_\pi \cdot L$ compared to devices based on LNOQ and undercut etched substrates due to a thinner silica buffer layer. For mass production, the structures can also be defined using deep UV lithography. The velocity matching is obtained in our proposed modulator based on a 12-μm-thick silica buried layer, which is favored in volume fabrication. Moreover, our device exhibits a lower cost, thanks to the relatively uncomplicated fabrication processes and their higher efficiency. Considering the balance between all the performance merits, our device has demonstrated an overall superior performance.

Table 1. Comparison of performance metrics for TFLN EO MZMs.

| Electrode (Substrate) | $V_\pi \cdot L$ (V·cm) | Length (mm) | EO roll-off | ER (dB) | Ref. |
|---|---|---|---|---|---|
| CPW (Si) | 2.8 | 20 | ~3 dB at 45 GHz | 30 | [10] |
| CPW (Si) | 1.29 | 3 | <3dB at 40 GHz | 19 | [16] |
| CPW (Si) | 1.02 | 5 | 3 dB at 108 GHz | / | [13] |
| CPW (hybrid) | 2.55 | 5 | ~3 dB at 70 GHz | / | [17] |
| Homologous CLTW (hybrid) | 3.1 | 5 | ~3 dB at 110 GHz | 28 | [18] |
| CLTW (quartz) | 2.3 | 10 | 1.8 dB at 50 GHz | 20 | [15] |
| CLTW (undercut etched Si) | 2.2 | 10 | ~1.4 dB at 67 GHz | >20 | [41] |
| CLTW (undercut etched Si) | 6.2 | 20 | 3 dB at 50 GHz | 24 | [42] |
| CLTW (undercut etched Si) | 2.125 | 12.5 | ~2 dB at 60 GHz | 25 | [46] |
| **This work** | **1.25** | **7** | **1.3 dB at 67 GHz** | **~30** | **-** |

## 4. Conclusion

In this work, we have proposed and demonstrated the CLTW electrode-based TFLN EO MZMs utilizing thick silica buffering layer. Experimentally, the device exhibits an ultra-flat EO frequency response and a high modulation efficiency, with a measured 1.3 dB roll-off at 67 GHz and a half-wave voltage length product of 1.25 V·cm in the telecom C-band. Moreover, we have proposed and demonstrated an efficient method to manage the microwave index, enabling perfect group velocity matching for both CLTW electrodes based on LNOQ and those based on thick-silica LNOI. Combined with the significantly reduced microwave loss afforded by the CLTW electrode, this represents the first demonstration of a TFLN modulator utilizing thick-silica LNOI that features ultra-high performance. We firmly believe that the TFLN platform based on a thick-silica buried layer, along with the proposed method for managing the microwave index, will serve as a cornerstone of research and application for high-performance TFLN modulators, thereby promoting the development of next-generation PICs.

**Acknowledgements**

**Author contributions**

X.X., Y.Z. and X.C. initiated the idea. X.X. and Y.X. performed the theoretical simulation. X.X., S.L. and H.L. designed and fabricated the device. X.X. conducted the experiment. W.D., R.Y., J.Q., S.L. and H.L. helped with the sample fabrication and the experiment. G.L., L.Y. and B.W. discussed the results. X.X. and Y.Z. wrote the paper with contribution from all the authors. Y.Z. and X.C. supervised the project.

**Conflict of interest**

The authors declare no competing interests.